\documentclass[acmtog]{acmart}
\usepackage{booktabs} 
\usepackage[normalem]{ulem}
\usepackage{algorithm} 
\usepackage{algorithmicx}
\usepackage[noend]{algpseudocode}
\usepackage{wrapfig}
\usepackage{comment}
\usepackage[final]{pdfpages}

\setcitestyle{square}

\begin{document}

\renewcommand\footnotetextcopyrightpermission[1]{}
\settopmatter{printacmref=false}

\title{Inverse Discrete Elastic Rod}

\author{Jiahao Li}
\email{lijh0417@mail.ustc.edu.cn}

\affiliation{%
  \institution{University of Science and Technology of China}
  \city{Hefei}
  \state{Anhui}
  \country{CN}
}

\author{Mingchao Liu}
\email{m.liu.2@bham.ac.uk}

\affiliation{%
  \institution{University of Birmingham}
  \city{Birmingham}
  \country{UK}
}

\author{Haiyi Liang}
\email{hyliang@ustc.edu.cn}

\affiliation{%
  \institution{University of Science and Technology of China}
  \city{Hefei}
  \state{Anhui}
  \country{CN}
}

\author{HengAn Wu}
\email{wuha@ustc.edu.cn}
\authornote{Corresponding author}

\affiliation{%
  \institution{University of Science and Technology of China}
  \city{Hefei}
  \state{Anhui}
  \country{CN}
}

\author{Weicheng Huang}
\email{weicheng.huang@newcastle.ac.uk}
\authornote{Corresponding author}

\affiliation{%
  \institution{Newcastle University}
  \city{Newcastle upon Tyne}
  \country{UK}
}

\renewcommand{\shortauthors}{Li et al.}

\begin{abstract}
Inverse design of slender elastic structures underlies a wide range of applications in computer graphics, flexible electronics, biomedical devices, and soft robotics.
Traditional optimization-based approaches, however, are often orders of magnitude slower than forward dynamic simulations and typically impose restrictive boundary conditions.
In this work, we present an inverse discrete elastic rods (inverse-DER) method that enables efficient and accurate inverse design under general loading and boundary conditions.
By reformulating the inverse problem as a static equilibrium in the reference configuration, our method attains computational efficiency comparable to forward simulations while preserving high fidelity.
This framework allows rapid determination of undeformed geometries for elastic fabrication structures that naturally deform into desired target shapes upon actuation or loading.
We validate the approach through both physical prototypes and forward simulations, demonstrating its accuracy, robustness, and potential for real-world design applications.

\end{abstract}



\begin{teaserfigure}  
\centering
\includegraphics[width=\textwidth]{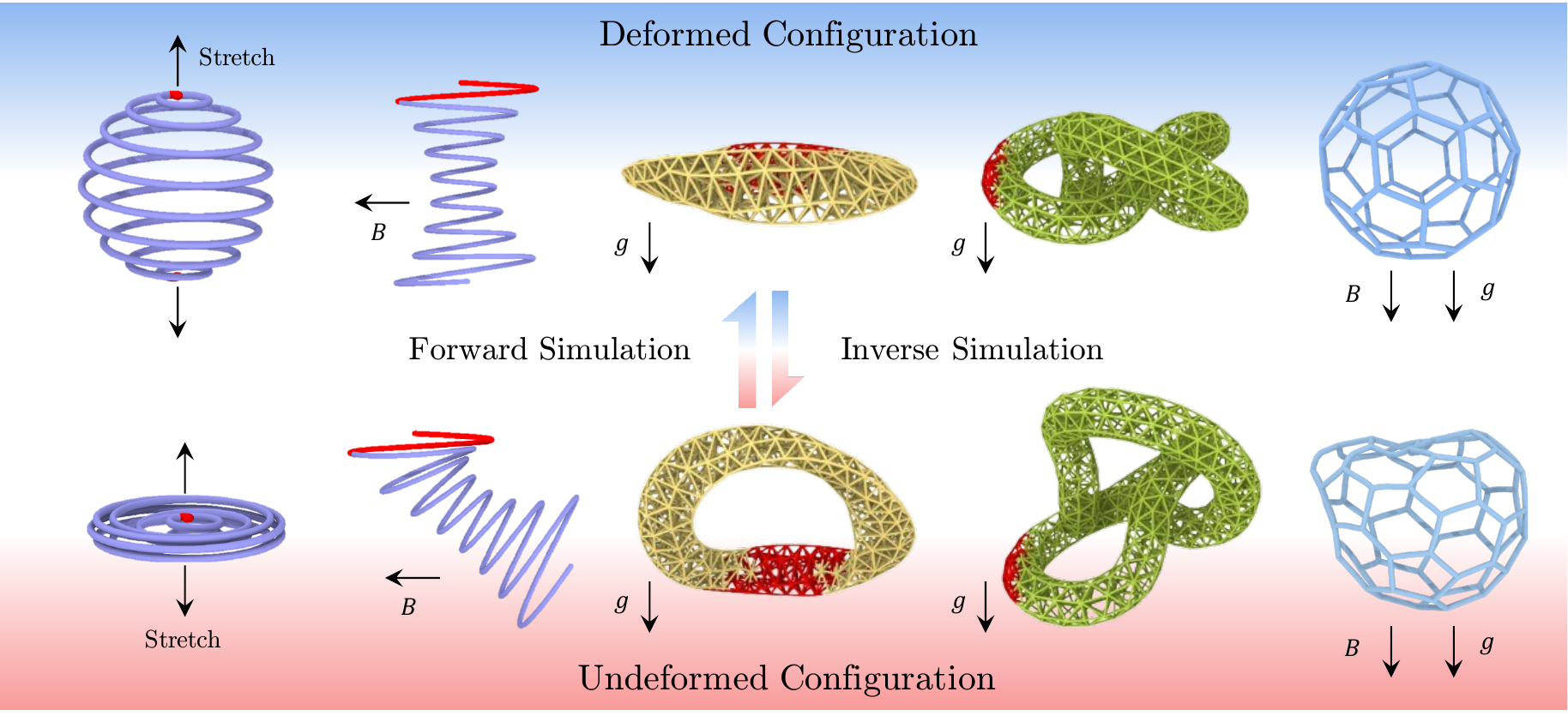}
\caption{This paper presents an inverse simulation framework for finding the undeformed configuration of a slender structure under arbitrary loading conditions. }
\label{fig:case6}
\end{teaserfigure}

\maketitle

\section{Introduction}

The simulation of elastic deformation is widely utilized in computer graphics~\cite{derouet2010stable,derouet2013inverse,ly2018inverse,bertails2018inverse,chen2014asymptotic}, engineering science~\cite{fan2020inverse,zhang2022shape,cheng2023programming}, and solid mechanics~\cite{liu2020tapered,yang2023morphing,yang2024hierarchical,li2025inverse,li2025biomimetic,tong2025inverse,huang2025tutorial}, and is increasingly important in emerging fields such as 3D printing and soft robotics.
Compared with three-dimensional (3D) solid structures, one-dimensional (1D) slender structures such as beams, rod, and ribbons are more versatile and flexible, thereby exhibiting large elastic deformation behavior~\cite{o2017modeling,audoly2000elasticity,langer1996lagrangian}.
Although the forward problem, solving the elastic deformation with boundary conditions to get the deformed configuration (DC), has been extensively studied over the past two decades~\cite{audoly2000elasticity,o2017modeling,love1944treatise,antman2005nonlinear,bergou2008discrete,bergou2010discrete,huang2025tutorial}, the inverse design problem, which seeks the undeformed configuration (UC) that produces a desired target shape under specific external loading conditions, remains a significant challenge. 

Recently, a number of studies have focused on the inverse design problem: finding the UC of a target shape~\cite{derouet2010stable,derouet2013inverse,miller2014shapes,bertails2018inverse,ly2018inverse,tong2025inverse,li2025biomimetic,li2025inverse}.
Current approaches for inverse design can be broadly divided into two categories: numerical optimization and theoretical solutions.
Inverse design is commonly formulated as a constrained minimization problem, owing to the absence of a comprehensive theoretical framework.
The objective is to minimize the discrepancy between the forward simulation result and a target shape, with applications spanning animation control of 2D curves, 3D curves~\cite{derouet2010stable, derouet2013inverse}.
However, when the problem is highly nonlinear and the initial guess is far from the true solution, these optimization-based methods tend to converge slowly and require substantial computational resources~\cite{chen2014asymptotic,derouet2013inverse,ly2018inverse}.
For example, utilizing a Lev-Mar solver for inverse design incurs a computational cost nearly two orders of magnitude higher than that of a forward simulation~\cite{chen2014asymptotic}.
In contrast, theoretical solutions are generally more efficient but are limited to relatively simple cases, e.g., a single rod system.
Bertail et al. established a theoretical foundation for the inverse design of gravity actuated suspended Kirchhoff rods; however, their framework is restricted to clamped-free boundary conditions~\cite{bertails2018inverse}.
Building on this, a unified theory, termed inverse elastica, was later developed, incorporating the geometric equations of the UC as presented in our previous work~\cite{li2025inverse}.
Although theoretical frameworks for the inverse design of slender structures exist, they struggle with geometrically complex systems such as nets, gridshells, and lattice-like systems~\cite{derouet2010stable,derouet2013inverse,bertails2018inverse,li2025inverse}.
Consequently, a general, optimization-free numerical method that leverages a unified theory is highly desired to efficiently tackle these challenging inverse design problems.

In this paper, we introduce the inverse discrete elastic rod (inverse-DER) method, built upon the discrete elastic rod (DER) framework, to address the inverse design of slender structures.
The DER method is widely adopted for simulating the large-deformation dynamics of slender bodies~\cite{bergou2008discrete,bergou2010discrete,huang2025tutorial}, capturing their evolution from a UC to a DC.
In a similar spirit, the proposed inverse-DER method enables the direct reconstruction of the UC through an inverse dynamic evolution process starting from a given DC.
In contrast to prior asymptotic approaches~\cite{chen2014asymptotic}, our framework eliminates the need for intricate mathematical derivations of asymptotic expansion coefficients.
For scenarios involving substantial geometric differences between the UC and DC, convergence is guaranteed by the analytic derivation of the Jacobian matrix for the inverse dynamics. 
Moreover, the computational cost of the inverse-DER method is on par with that of a standard forward simulation.
Contrary to previous optimization-based strategies~\cite{derouet2010stable,derouet2013inverse,ly2018inverse}, which are typically limited to clamped-free boundary conditions and simple loading conditions, our approach covers more general boundary conditions (BCs) such as clamped-free and clamped-clamped BCs, as well as general loading conditions, e.g., gravitational and magnetic actuations.
Using our method, we achieve the inverse design of rods under general BCs, as well as more complex networks subjected to gravity and magnetic fields.
Our results are validated through both experimental fabrications and forward simulations, demonstrating the high performance and robustness of the inverse-DER framework.
In summary, the main contributions of this work include:

\begin{itemize}

    \item \textbf{Inverse Simulation Framework} We propose a general inverse simulation framework by formulating force equilibrium in reference configuration rather than current configuration.
    The computational efficiency of the inverse simulation is comparable to that of the forward simulation.
  
    \item \textbf{Inverse Discrete Elastic Rod} Building upon the classical DER formulation, we develop inverse-DER to reconstruct the undeformed configuration (UC) from the deformed configuration (DC) via direct inverse dynamics.
    In contrast to traditional DER, our method explicitly incorporates parallel transport between the reference and current configurations when computing the force and Jacobian, thereby ensuring path independence.

    \item \textbf{Nets Extension and Validation}  We achieve inverse design of curve-discretized surfaces with various Gaussian curvatures under different BCs and loading conditions.
    We then extend the framework to nets, enabling the inverse design of twisting ring nets and trefoil knot nets under the actuation of gravity.
    Finally, we investigate inverse design of fullerene networks under both gravity and magnetic fields.

    \item \textbf{Computational Efficiency, Solution Existence and Energy Profiles} We compare the computational efficiency between DER and inverse-DER to ensure the accuracy and correctness of our inverse simulation framework.
    The existence of a solution can be directly obtained by checking for the singularity of the local Jacobian matrix.
    Furthermore, by comparing the energy profiles during forward and inverse simulation, we elucidate the fundamental distinction between inverse simulation and time-reversal symmetry.
    
\end{itemize}

\section{Related Work}

The inverse design of slender structures has garnered significant interest across numerous fields, such as computer graphics~\cite{derouet2010stable,derouet2013inverse,chen2014asymptotic,ly2018inverse,bertails2018inverse}, solid mechanics~\cite{miller2014shapes,li2025biomimetic,li2025inverse,tong2025inverse}, and flexible electronics~\cite{fan2020inverse, cheng2023programming}.
Despite considerable advances in theoretical and computational methodologies, the inverse design of complex structures remains a substantial challenge.
This section reviews recent, highly relevant work on inverse design strategies.
These strategies can be broadly categorized into two paradigms: theoretical methods and numerical algorithm.

\paragraph{Inverse Design Algorithm} A wide range of numerical algorithms have been developed to address the inverse problems.
The core idea behind numerical approaches is to solve the forward equations using an initially guessed configuration and then iteratively minimize the discrepancy between the computed solution and the target shape.
For instance, Hadap computed zero-gravity rest shapes of strands using a multi-body reduced model~\cite{hadap2006oriented}.
Derouet-Jourdan et al. introduced an inverse design technique for 2D dynamic curves that converts user-drawn sketches into dynamic rod models and optimizes natural curvatures to achieve specific equilibrium states~\cite{derouet2010stable}.
Subsequently, Derouet-Jourdan et al. proposed a constrained optimization method to solve the inverse static equilibrium problem for hairs under gravity and frictional contact~\cite{derouet2013inverse}.
As a follow-up work, they extended this method to shell models ~\cite{ly2018inverse}.
Although these studies successfully demonstrated inverse design, their reliance on optimization techniques tends to result in high computational cost and substantial resource demands.
To improve efficiency, Chen et al. developed an algorithm based on the asymptotic numerical method~\cite{chen2014asymptotic}.
They compute the numerical asymptotic parameters to increase the external loading gradually and successfully recovering the UC with neo-Hookean materials from DC under gravity.
However, this method requires sophisticated mathematical derivation of asymptotic expansion coefficients and remains limited to objects with clamped-free boundary conditions under external loading.

\paragraph{Inverse Design Theory} Compared with numerical approaches, theoretical methods are generally more computationally efficient.
Nevertheless, theoretical studies on the inverse design of slender structures remain relatively limited.
Bertail et al. developed an inverse theory to determine the UC of an isotropic suspended Kirchhoff rod under gravity, but their analysis was restricted to rods with circular cross-sections and clamped-free boundary conditions~\cite{bertails2018inverse}.
As a unified framework, Li et al. developed inverse elastica to address the inverse design of slender structures under general boundary conditions~\cite{li2025inverse}.
Within this framework, they proposed the concept of inverse loading, which enables direct reconstruction of UC from DC. Despite the establishment of inverse elastica as a general theoretical foundation, solution of the inverse elastica ordinary differential equations remains applicable only to simple cases.
For complex geometries composed of slender structures, such as nets, theoretical solutions become practically infeasible.
Therefore, the development of a general numerical method is both necessary and urgent.

\paragraph{Discrete Elastic Rod} The Discrete Elastic Rods (DER) method has been established as a highly efficient framework for simulating the dynamics of slender structures~\cite{bergou2008discrete,bergou2010discrete,huang2025tutorial}, with broad applications in engineering mechanics~\cite{huang2020shear,tong2025inverse}, computer graphics~\cite{bertails2018inverse,bergou2008discrete}, and soft robotics~\cite{li2025harnessing,tong2025real}.
In DER, force equilibrium in the current configuration is enforced by solving the governing equations of motion.
Two features are particularly notable.
First, the centerline is represented explicitly, which facilitates the simulation of complex contact and looping phenomena.
Second, the twisting strain is formulated as the rotation angle between the material frame and reference frame, with the latter updated through parallel transport.  
For inverse-DER, we also use the reference frame to evaluate twisting forces, but with two important differences from DER.
The reference frame at time step $t_k$ is not propagated from the previous frame at $t_{k-1}$, but instead obtained directly by parallel transport from the initial frame at $t_0$, thereby avoiding the storage of intermediate frames, and the final result is path-dependent.
In addition, forces are evaluated in the reference configuration rather than the current configuration.
These differences will be discussed in detail in Section~4.

\section{Overview}

We first introduce the inverse simulation framework by formulating an inverse dynamics process in Section 4.1.
Next in Section 4.2, we build the inverse discrete elastic rod method framework with the proposed inverse simulation framework applied. The difference between DER and inverse-DER is also investigated in this part.
In Section 5, we demonstrate the method’s applicability through a series of inverse design cases, including both slender and net structures.
Section 6 further discusses the computational efficiency, solution existence, and energy profiles of the proposed inverse-DER approach, and outlines potential directions for future work.
The paper concludes with a summary in Section 7.

\begin{figure}[h]  
\centering
\includegraphics[width=\linewidth]{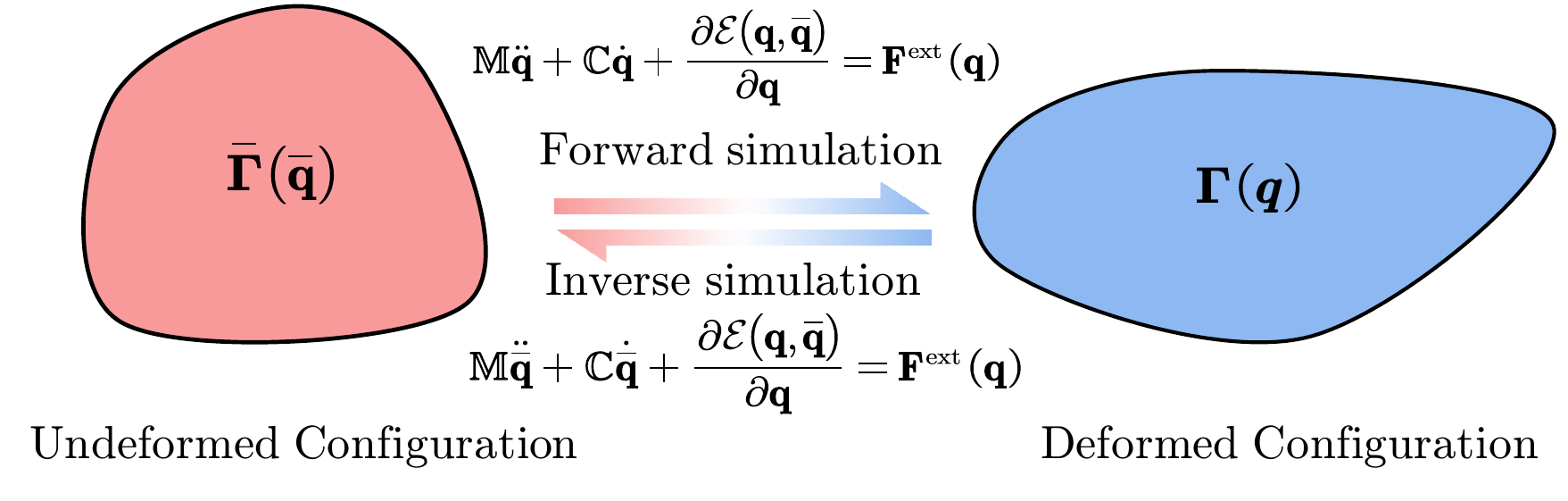}
\caption{The forward and inverse simulation framework.}
\label{fig:frame}
\end{figure}

\section{Numerical method}

Here, we first present the forward and inverse simulation framework for a general nonlinear structural system, and then introduce the numerical framework for the slender rod system.

\subsection{General Forward and Inverse Simulation Framework}

To facilitate understanding, we provide an illustrative comparison between forward and inverse simulation. 
As shown in Fig. ~\ref{fig:frame}, we have two configurations, DC and UC.
We first consider the forward simulation from the red UC to the blue DC in Fig. ~\ref{fig:frame}.
Here, the general degrees of freedom (DOF) vector of the dynamic system is denoted as $\mathbf{q}$, and $\bar{\mathbf{q}}$ represents the initial DOF vector.
Hereafter, a bar on top indicates the evaluation on the UC.
For forward simulation, we take UC as the reference configuration $\bar{\mathbf{\Gamma}}( \bar{\mathbf{q}} )$, and take DC as the current configuration $\mathbf{\Gamma}(\mathbf{q})$; we aim to solve the dynamic equations of motion in the current configuration.
Therefore, the forward dynamic equation can be written as:
\begin{equation}
\mathbb{M}\ddot{\mathbf{q}}+\mathbb{C}\dot{\mathbf{q}}+\frac{\partial\mathcal{E}(\mathbf{q}, \bar{\mathbf{q}})}{\partial\mathbf{q}}=\mathbf{F}^{\mathrm{ext}}(\mathbf{q}),
\label{eqn:for}
\end{equation}
where $\mathbb{M}$ is the mass matrix, $\mathbb{C}$ is the damping matrix, $\mathcal{E}$ is the total potential energy, and $\mathbf{F}^{\mathrm{ext}}$ is the external force vector.
Usually, the implicit Euler method is employed to numerically solve the dynamic equilibrium equation step  by step, and the static equilibrium configuration can be derived through the dynamic relaxation method, i.e., we can get the solutions for
\begin{equation}
\frac{\partial\mathcal{E}(\mathbf{q}, \bar{\mathbf{q}})}{\partial\mathbf{q}}=\mathbf{F}^{\mathrm{ext}}(\mathbf{q})
\end{equation}
Overall, for the forward simulation, we take the UC as the reference configuration, and use Eq.~\eqref{eqn:for} to solve the forward dynamic evolution from UC to DC.

On the other side, for inverse problem, DC is known, and we aim to design UC based on the loading and boundary conditions (BCs).
Therefore, we aim to design a corresponding inverse dynamic evolution from DC to UC.
Instead of an optimization-based approach,  we here propose an inverse simulation framework.
In the inverse simulation, we take DC, $\mathbf{\Gamma}(\mathbf{q})$, as the reference configuration , and take UC, $\bar{\mathbf{\Gamma}}(\bar{\mathbf{q}})$, as the current configuration.
We need to solve $\bar{\mathbf{q}}$ from the equilibrium equation in reference configuration $\mathbf{\Gamma}(\mathbf{q})$, and, therefore, the inverse dynamic equation can be written as:
\begin{equation}
\mathbb{M} \ddot{\bar{\mathbf{q}}} + \mathbb{C}\dot{\bar{\mathbf{q}}} + \frac{\partial\mathcal{E}(\mathbf{q}, \bar{\mathbf{q}})}{\partial\mathbf{q}}=\mathbf{F}^{\mathrm{ext}}(\mathbf{q}).
\label{eqn:inv}
\end{equation}
Similarly, we take the DC as the reference configuration, and the final static solution, $\bar{\mathbf{q}}$, for inverse design can be derived through the dynamic relaxation method.

For a clear comparison, Table~\ref{T1} summarizes the key differences between inverse and forward simulation. In the forward simulation, the UC is the reference configuration, while the DC is the current configuration; the force equilibrium is solved in the current configuration. Conversely, in the inverse simulation, the DC is taken as the reference configuration, and the UC is the current configuration; the equilibrium is solved in the reference configuration.

\begin{table*}[t]
\centering
\normalsize
\begin{tabular}{llll}
\hline
 & Reference configuration  & Current configuration   & Force equilibrium \\ \hline
Forward simulation & Undeformed configuration & Deformed configuration   & Current configuration   \\ \hline
Inverse simulation & Deformed configuration   & Undeformed configuration & Reference configuration  \\ \hline
\end{tabular}
\caption{Comparison between forward and inverse simulations.}
\label{T1}
\end{table*}

\subsection{Forward and Inverse Discrete Elastic Rods}

\begin{figure}[h]  
\centering
\includegraphics[width=\linewidth]{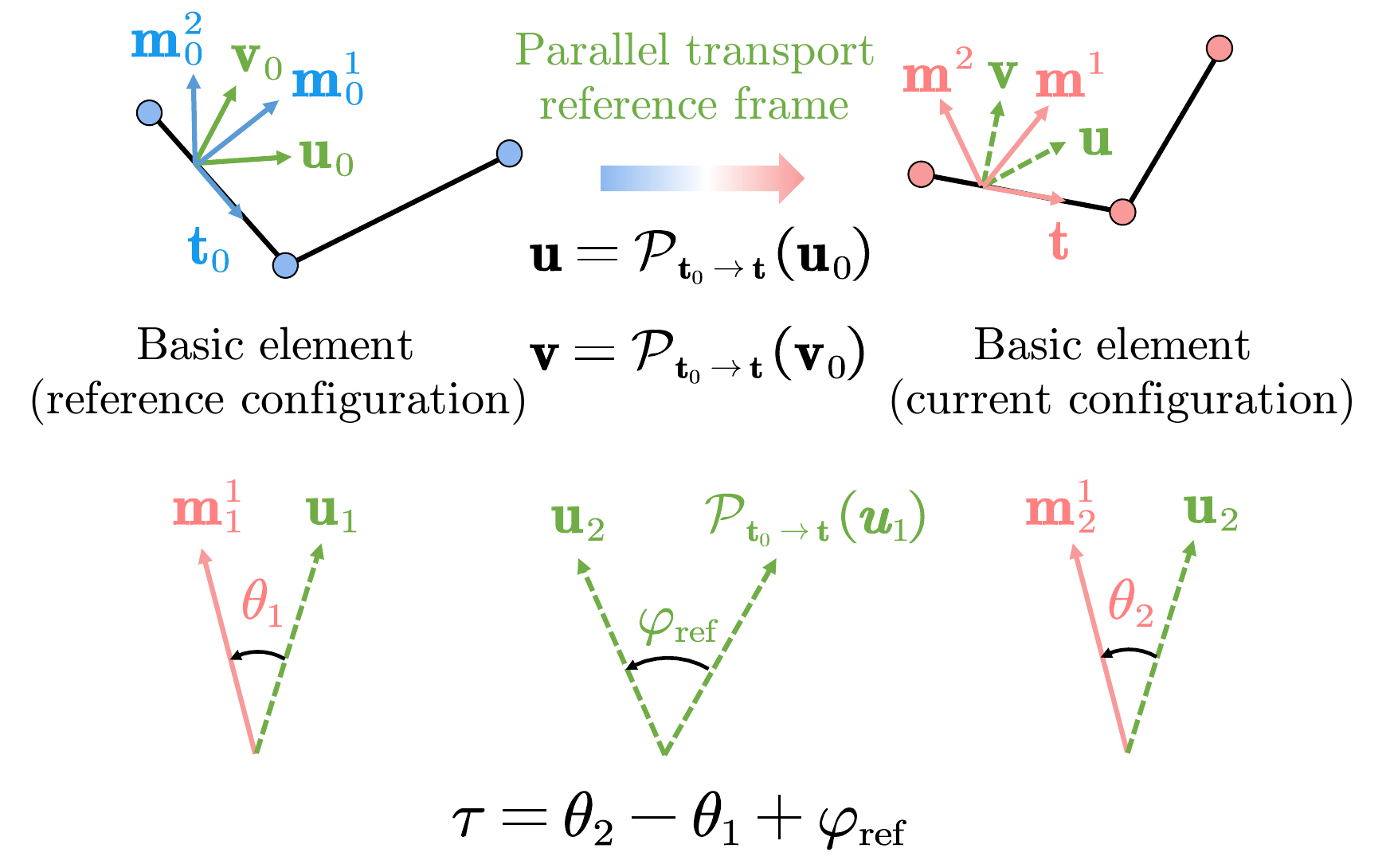}
\caption{\textbf{The computational detail of the discrete 3D curve.} (a) The parallel transport between the reference configuration and the current configuration. (b) The computation of the discrete reference twist.}
\label{fig:force}
\end{figure}

In this subsection, we present the numerical method for both forward and inverse-DER simulation.
We first review the elastic energy formulation in a discrete rod system, then focus on forward and inverse solvers.

\paragraph{Rod model} We use a Kirchhoff model to formulate the mechanics of a slender system with $N_v$ nodes and $N_e$ edges.
Each edge is defined as a stretching element, and a bending element is defined as a pair of adjacent edges, resulting in $N_b$ bending elements. 
Two frames are constructed at each edge: (i) an adaptive frame and (ii) a material frame.
As shown in Fig.~\ref{fig:force}(a), we use $\{\mathbf{u}, \mathbf{v}, \mathbf{t} \}$ define the adaptive frame for each  edge, and use $\{\mathbf{m}^{1},\mathbf{m}^{2}, \mathbf{m}^{3}  \}$ to define the its material frame.
To achieve a path-independent frame updating from the initial reference configuration to the current configuration, the parallel transport method is employed. 
An operator, $\mathcal{P}$, is used to define the parallel transport,
\begin{equation}
\mathbf{a}_{2}  = \mathcal{P}_{\mathbf{t}_{1} \rightarrow \mathbf{t}_{2} }(\mathbf{a}_{1}),
\end{equation}
which means that when a tangential vector, $\mathbf{t}_{1}$, rotates to another direction, $\mathbf{t}_{2}$, the associated frame, $\mathbf{a}_{1}$, also can also rotate to, $\mathbf{a}_{2}$, without any twist.
Thus, with a given adaptive frame in the initial reference configuration, $\{\mathbf{u}_{0}, \mathbf{v}_{0}, \mathbf{t}_{0}\}$, the adaptive frame in the current configuration is 
\begin{equation}
\begin{cases}
\mathbf{u} = \mathcal{P}_{\mathbf{t}_{0} \rightarrow \mathbf{t} }(\mathbf{u}_{0}), \\
\mathbf{v} =\mathcal{P}_{\mathbf{t}_{0} \rightarrow \mathbf{t} }(\mathbf{v}_{0}).
\end{cases}
\end{equation}
Next, the material frame of the current configuration is updated by the rotation angle and reference frame:
\begin{equation}
\begin{cases}
\mathbf{m}^{1}= \mathbf{u} \cos \theta +\mathbf{v}\sin \theta,  \\
\mathbf{m}^{2}=\mathbf{v}\cos \theta -\mathbf{u} \sin \theta, \\
\mathbf{m}^{3}= \mathbf{t}.
\end{cases}
\end{equation}
Finally, the total DOF vector is the sum of nodal positions and twisting angle, 
$\mathbf{q} \in  \mathcal{R}^{(3N_{v}+N_{e}) \times 1}$.
For consistency, we use a bar on top, $\bar{()}$, to define the variables in the UC, i.e., $\bar{\mathbf{q}}$ is the DOF vector in the stress-free UC. 

The total elastic energy of a discrete slender system is comprised of three components:  stretching, bending, and twisting.
The stretching element is comprised of two nodes, i.e.,  the $i$-th stretching element is $ \mathcal{S}: \{ \mathbf{x}_{1}, \mathbf{x}_{2}\}$, and we use $\mathbf{e}$ to denote the edge vector between the two nodes, thus uniaxial stretching strain can be obtained as
\begin{equation}
\epsilon = \frac{\| \mathbf{e} \|}{\| \bar{\mathbf{e}} \|} -1,
\end{equation} 
and the elastic stretching energy is the sum over all stretching elements,
\begin{equation}
\mathcal{E}_s = \frac {1} {2} \sum_{i=0}^{N_{e}} EA (\epsilon_{i})^2 || \bar{\mathbf{e}}_{i} || 
\end{equation} 
where  $EA$  is the stretching stiffness.

\begin{figure}[h]  
\centering
\includegraphics[width=\linewidth]{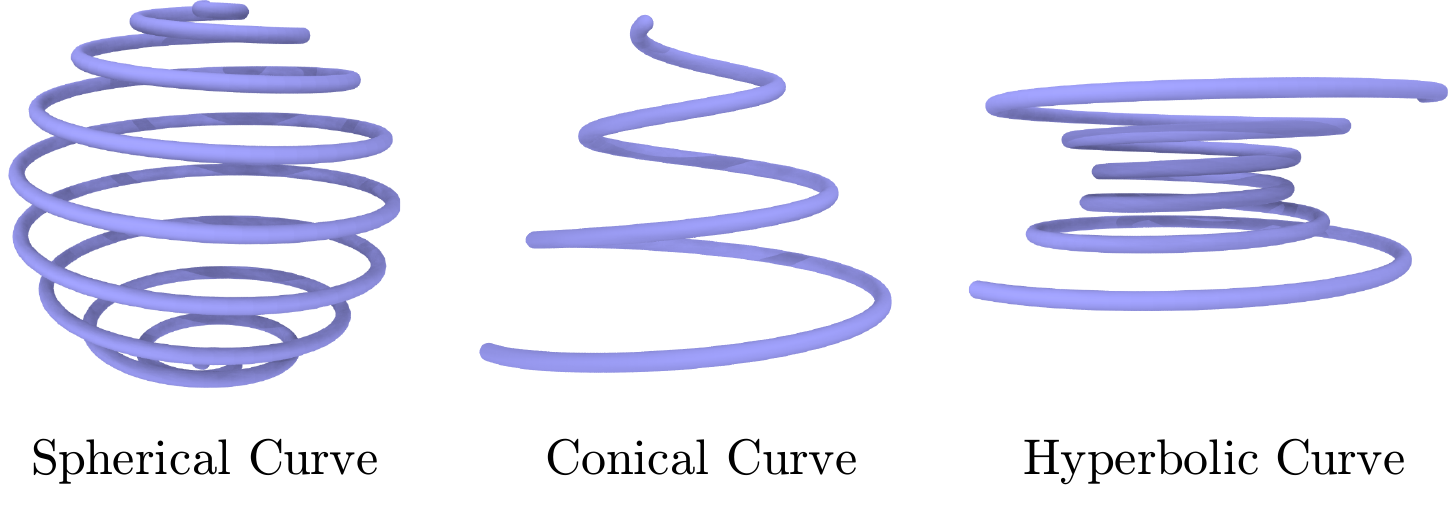}
\caption{\textbf{The three curve-discretized surfaces with various Gaussian curvatures.} (a) Spherical curve. (b) Conical curve. (c) Hyperbolic curve.}
\label{fig:case123}
\end{figure}

\begin{figure*}[h]  
\centering
\includegraphics[width=\textwidth]{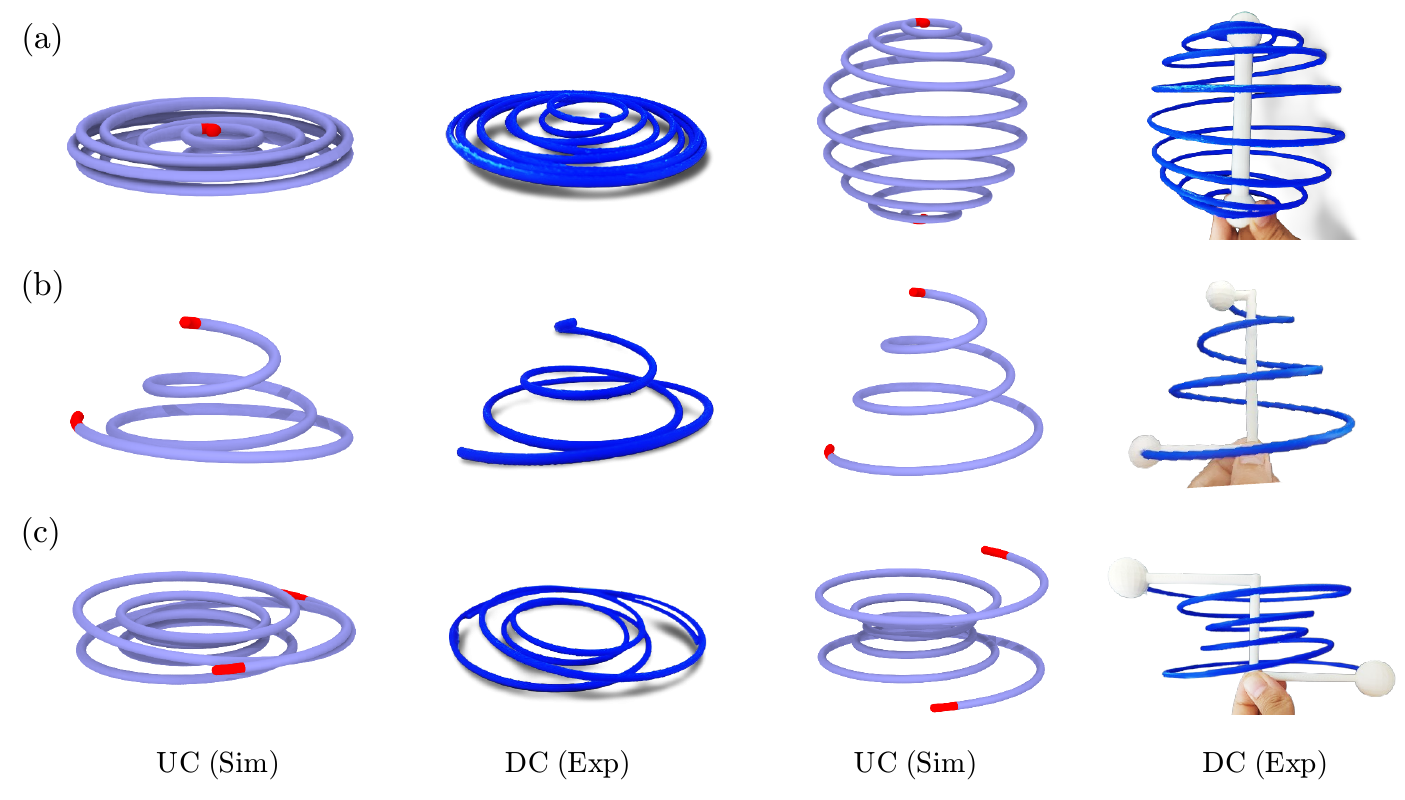}
\caption{\textbf{Inverse design of compressed curve-discritized surfaces.} (a) The comparison between simulation and experimental results of UC and DC for the spherical curve. (b) The comparison between simulation and experimental results of UC and DC for the conical curve. (c) The comparison between simulation and experimental results of UC and DC for the hyperbolic curve. The red parts of two ends are clamped.}
\label{fig:case1}
\end{figure*}

The bending element is comprised of three consecutive nodes and two adjacent edges  $ \mathcal{B}: \{ \mathbf{x}_{1}, \theta_{1} \mathbf{x}_{2}, \theta_{2}, \mathbf{x}_{3} \}$.
Similarly, we use $\mathbf{e}_{1}$ and $\mathbf{e}_{2}$ to denote the two edge vectors, and use the  $\Delta l$ to denote its Voronoi length, which is the average of the two edges.
The material frames for two edges are $\{\mathbf{m}_{1}^{1},\mathbf{m}_{1}^{2}, \mathbf{m}_{1}^{3}  \}$ and $\{\mathbf{m}_{2}^{1},\mathbf{m}_{2}^{2}, \mathbf{m}_{2}^{3}  \}$.
The associated bending strain is captured by the curvature binormal, which measures the misalignment between two consecutive edges,
\begin{equation}
\mathbf{\kappa b} = \frac {2 (\mathbf{e}_{1} \times \mathbf{e}_{2}) } { \| \mathbf{e}_{1} \| \| \mathbf{e}_{2} \| + \mathbf{e}_{1} \cdot \mathbf{e}_{2} },
\end{equation}
and the inner products between the curvature binormal and material frame vectors give the material curvatures,
\begin{equation}
\begin{aligned}
\kappa_{1} & = \frac{1}{2} \left( \mathbf m_1^{1} + \mathbf m_2^{1} \right) \cdot \kappa \mathbf b, \\
\kappa_{2} & = - \frac{1}{2} \left( \mathbf m_1^{2} + \mathbf m_2^{2} \right) \cdot \kappa \mathbf b.
\end{aligned}
\label{eq:bendingStrain}
\end{equation}
The total elastic bending energy is 
\begin{equation}
\mathcal{E}_b = \frac {1} {2} \sum_{i=0}^{N_{b}} \left[ \frac {EI_{1}} { \Delta {l}_{i} }  (\kappa_{1,i} - \bar{\kappa}_{1,i})^2 +   \frac {EI_{2}} { \Delta {l}_{i} } (\kappa_{2,i} - \bar{\kappa}_{2,i})^2 \right],
\end{equation}
where $EI_{1}$  and  $EI_{2}$  are the local bending stiffness along two material directions.

\begin{figure*}[h]  
\centering
\includegraphics[width=\textwidth]{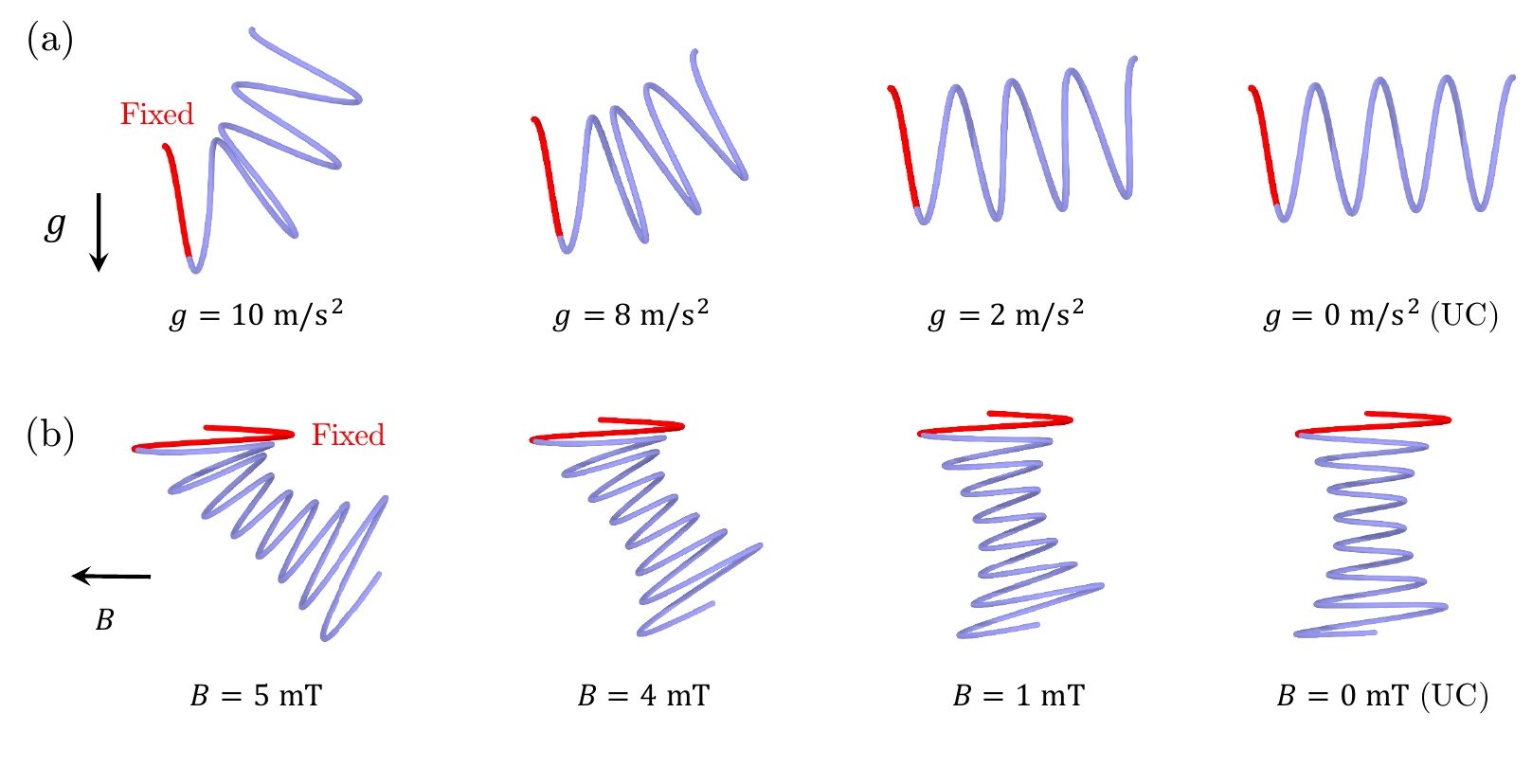}
\caption{\textbf{Inverse design of helical curve under gravity and hyperbolic curve under magnetic field.} (a) Undeformed shapes of helical curves subjected to varying gravitational strengths. (b) Undeformed shapes of hyperbolic curves subjected to different magnetic field intensities.}
\label{fig:case23}
\end{figure*}

The twisting element is the same as the bending element,  $ \mathcal{B}: \{ \mathbf{x}_{1}, \theta_{1} \mathbf{x}_{2}, \theta_{2}, \mathbf{x}_{3} \}$, and, as shown in Fig.~\ref{fig:force}(b), the twisting strain can be computed as 
\begin{equation}
\tau =  \theta_{2}-\theta_{1} +\varphi_{\mathrm{ref}} , 
\end{equation}
where $\varphi_{\mathrm{ref}}$ is the reference twist and can be computed as:
\begin{equation}
\varphi_{\mathrm{ref}} = \measuredangle \Bigl(\mathcal{P}_{\mathbf{t}_{1} \rightarrow \mathbf{t}_{2} }( \mathbf{u}_1), \mathbf{u}_{2},  \Bigl),
\end{equation}
where $\measuredangle (\mathbf{a},\mathbf{b})$ means the relative rotational angle of between two vectors, $\{ \mathbf{a}, \mathbf{b} \}$, along their binormal direction, $\mathbf{a} \times \mathbf{b}$.
Note that the reference frame directors, $\{\mathbf{u}_{1}, \mathbf{u}_{2} \}$, are updated based on the initial reference configuration by using the parallel transport, instead of the previous time step, to avoid the path-dependency. 
The twisting energy is also the quadratic form of the twisting strain,
\begin{equation}
\mathcal{E}_t = \frac {1} {2} \sum_{i=0}^{N_{t}} \frac {GJ} { \Delta {l}_{i} } ( \tau_{i} - \bar{\tau}_{i} )^2,
\end{equation}
where $GJ$ is the local twisting stiffness.

The geometry of a single rod is straightforward and well-defined.
However, when considering arbitrary configurations of complex net-like structures, the sign of the twisting angle must be adjusted: if the directions of the two edges are opposite, the twisting direction of the angle will also reverse.
By introducing this modification, our framework becomes applicable to slender structures with diverse topologies.

\begin{figure*}[h]  
\centering
\includegraphics[width=\linewidth]{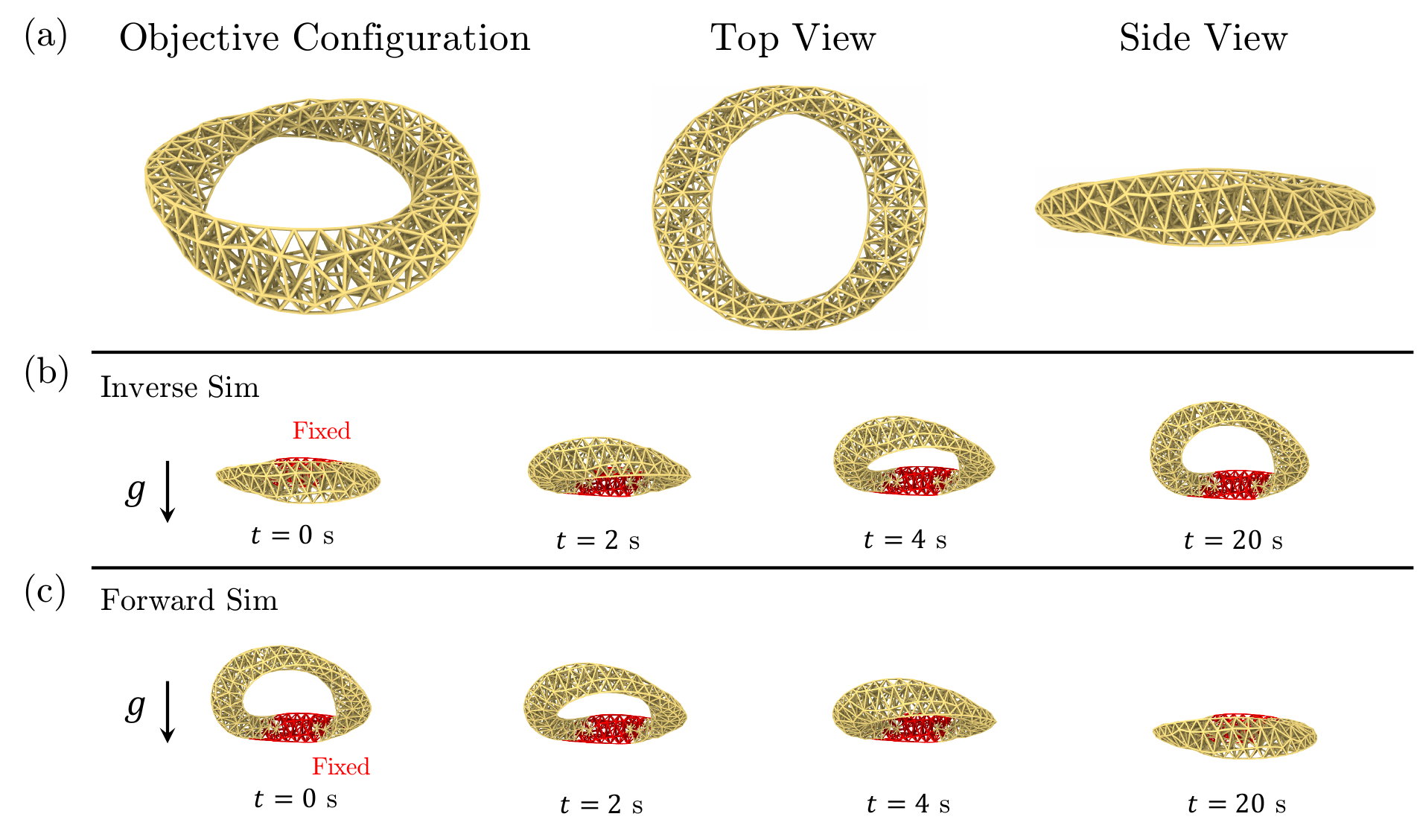}
\caption{\textbf{The inverse and forward simulation of net ring.} (a) The objective net ring. (b) The inverse simulation of net ring. (c) The verification of the inverse simulation results by forward simulation.}
\label{fig:case4}
\end{figure*}

\paragraph{Forward simulation}  In the forward DER, the UC is known and is treated as the reference configuration, $\mathbf{q}(t=0) = \bar{\mathbf{q}}$, and is denoted as ${\mathbf{q}}_{0}$; we need to solve for $\mathbf{q}(t)$ at each time step.
The residual error of force can be written as: 
\begin{equation}
\mathbf{R}^{\mathrm{for}}({\mathbf{q}}) \equiv \mathbb{M} \ddot{{\mathbf{q}}} + \mathbb{C}\dot{{\mathbf{q}}}+\frac{\partial (\mathcal{E}_{s} + \mathcal{E}_{b} + \mathcal{E}_{t})}{\partial\mathbf{q}} - \mathbf{F}^{\mathrm{ext}}(\mathbf{q}).
\end{equation}
where the elastic force vector is associated with the gradient of energy in the DC, $\mathbf{q}(t)$, 
\begin{equation}
\begin{aligned}
    \mathbf{F}_{s}^{\mathrm{for}}  &= - \frac{\partial \mathcal{E}_s}{\partial \mathbf{q}} =  -\sum_{i=0}^{N_{e}} \frac{\partial \mathcal{E} _s}{\partial \epsilon_{i}} \frac{\partial \epsilon_{i} }{\partial \mathbf{q}} \\
    \mathbf{F}_{b}^{\mathrm{for}}  &= - \frac{\partial \mathcal{E}_b}{\partial \mathbf{q}} =  -\sum_{i=0}^{N_{b}} \left( \frac{\partial \mathcal{E}_b }{\partial \kappa_{1,i}} \frac{\partial \kappa_{1,i} }{\partial \mathbf{q}} +  \frac{\partial \mathcal{E}_b }{\partial \kappa_{2,i}} \frac{\partial \kappa_{2,i} }{\partial \mathbf{q}}  \right) \\
    \mathbf{F}_{t}^{\mathrm{for}}  &= - \frac{\partial \mathcal{E}_t}{\partial \mathbf{q}} =  -\sum_{i=0}^{N_{b}} \frac{\partial \mathcal{E}_t}{\partial \tau_{i}} \frac{\partial \tau_{i} }{\partial \mathbf{q}}.
\end{aligned}
\end{equation}
Here, the first term (e.g., ${\partial \mathcal{E} _s} /{\partial \epsilon_{i}}$) is material-dependent and is simple to calculate, and the second term  (e.g.,  $ {\partial \epsilon_{i} } /{\partial \mathbf{q}}  $) is geometry-dependent and can be found in the original DER formulation.
In forward DER, we need to solve for $\mathbf{q}(t)$ at each time step, thus, the Hession matrix is given by, 
\begin{equation}
\begin{aligned}
    \mathbb{J}_{s}^{\mathrm{for}} = \frac{\partial} {\partial \mathbf{q}} \left( \frac{\partial \mathcal{E}_s}{\partial \mathbf{q}} \right) &=  \sum_{i=0}^{N_{e}} \left[ \frac{\partial^2 \mathcal{E} _s}{\partial \epsilon_{i}^2} \left(\frac{\partial \epsilon_{i} }{\partial \mathbf{q}} \otimes \frac{\partial \epsilon_{i} }{\partial \mathbf{q}} \right) + \frac{\partial \mathcal{E} _s}{\partial \epsilon_{i}} \frac{\partial^2 \epsilon_{i} }{\partial \mathbf{q} \partial \mathbf{q}}   \right] \\
    \mathbb{J}_{b}^{\mathrm{for}} = \frac{\partial} {\partial \mathbf{q}} \left( \frac{\partial \mathcal{E}_b}{\partial \mathbf{q}} \right)  &=  \sum_{i=0}^{N_{b}} \left[ \frac{\partial^2 \mathcal{E}_b }{\partial \kappa_{1,i}^2} \left( \frac{\partial \kappa_{1,i} }{\partial \mathbf{q}} \otimes  \frac{\partial \kappa_{1,i} }{\partial \mathbf{q}} \right) + \frac{\partial \mathcal{E}_b }{\partial \kappa_{1,i}} \frac{\partial \kappa_{1,i}^2 }{\partial \mathbf{q} \partial \mathbf{q}} \right]  \\
    & + \sum_{i=0}^{N_{b}}\left[ \frac{\partial^2 \mathcal{E}_b }{\partial \kappa_{2,i}^2} \left( \frac{\partial \kappa_{2,i} }{\partial \mathbf{q}} \otimes  \frac{\partial \kappa_{2,i} }{\partial \mathbf{q}} \right) + \frac{\partial \mathcal{E}_b }{\partial \kappa_{2,i}} \frac{\partial \kappa_{2,i}^2 }{\partial \mathbf{q} \partial \mathbf{q}} \right] \\
    \mathbb{J}_{t}^{\mathrm{for}} = \frac{\partial} {\partial \mathbf{q}} \left( \frac{\partial \mathcal{E}_t}{\partial \mathbf{q}} \right) &=  \sum_{i=0}^{N_{b}} \left[ \frac{\partial^2 \mathcal{E}_t}{\partial \tau_{i}^2} \left(\frac{\partial \tau_{i} }{\partial \mathbf{q}} \otimes \frac{\partial \tau_{i} }{\partial \mathbf{q}} \right) + \frac{\partial \mathcal{E}_t}{\partial \tau_{i}} \frac{\partial^2 \tau_{i} }{\partial \mathbf{q} \partial \mathbf{q}}   \right].
\end{aligned}
\end{equation}
Newton's iteration is used to optimize the DOF vector $\mathbf{q}(t)$ to ensure the error is smaller than the tolerance.

\paragraph{Inverse simulation} In the inverse-DER, the DC is known and is treated as the reference configuration, $\bar{\mathbf{q}}(t=0) = {\mathbf{q}}$, and is denoted as $\bar{\mathbf{q}}_{0}$; we need to solve for $\bar{\mathbf{q}}(t)$ at each time step.
The residual error of force can be written as: 
\begin{equation}
\mathbf{R}^{\mathrm{inv}}(\bar{{\mathbf{q}}}) \equiv \mathbb{M} \ddot{ \bar{\mathbf{q}}} + \mathbb{C}\dot{ \bar{\mathbf{q}}}+\frac{\partial (\mathcal{E}_{s} + \mathcal{E}_{b} + \mathcal{E}_{t})}{\partial\mathbf{q}} - \mathbf{F}^{\mathrm{ext}}(\mathbf{q}).
\end{equation}
where the elastic force vector is associated with the gradient of energy in the DC, $\mathbf{q}$, which is also the reference configuration, $\bar{\mathbf{q}}_0$, instead of the current UC, $\bar{\mathbf{q}}(t)$,
\begin{equation}
\begin{aligned}
    \mathbf{F}_{s}^{\mathrm{inv}}  &= - \frac{\partial \mathcal{E}_s}{\partial \mathbf{q}} =  -\sum_{i=0}^{N_{e}} \frac{\partial \mathcal{E} _s}{\partial \epsilon_{i}} \frac{\partial \epsilon_{i} }{\partial \bar{\mathbf{q}}_{0}} \\
    \mathbf{F}_{b}^{\mathrm{inv}}  &= - \frac{\partial \mathcal{E}_b}{\partial \mathbf{q}} =  -\sum_{i=0}^{N_{b}} \left( \frac{\partial \mathcal{E}_b }{\partial \kappa_{1,i}} \frac{\partial \kappa_{1,i} }{\partial \bar{\mathbf{q}}_{0}} +  \frac{\partial \mathcal{E}_b }{\partial \kappa_{2,i}} \frac{\partial \kappa_{2,i} }{\partial \bar{\mathbf{q}}_{0} }  \right) \\
    \mathbf{F}_{t}^{\mathrm{inv}}  &= - \frac{\partial \mathcal{E}_t}{\partial \mathbf{q}} =  -\sum_{i=0}^{N_{b}} \frac{\partial \mathcal{E}_t}{\partial \tau_{i}} \frac{\partial \tau_{i} }{\partial \bar{\mathbf{q}}_{0} }.
\end{aligned}
\end{equation}
Also, we need to solve for $\bar{\mathbf{q}}(t)$ at each time step, and, thus, the Hessian matrix is the gradient to the current configuration, 
\begin{equation}
\begin{aligned}
    \mathbb{J}_{s}^\mathrm{inv} = \frac{\partial} {\partial \bar{\mathbf{q}}} \left( \frac{\partial \mathcal{E}_s}{\partial \mathbf{q}} \right) &=  \sum_{i=0}^{N_{e}} \left[ \frac{\partial^2 \mathcal{E} _s}{\partial \bar{\epsilon}_{i} \partial \epsilon_{i}  } \left(\frac{\partial \epsilon_{i} }{\partial \bar{\mathbf{q}}_{0}  } \otimes \frac{\partial \epsilon_{i} }{\partial \bar{\mathbf{q}}} \right)  \right] \\
    \mathbb{J}_{b}^\mathrm{inv}  = \frac{\partial} {\partial \bar{\mathbf{q}}} \left( \frac{\partial \mathcal{E}_b}{\partial \mathbf{q}} \right) & =  \sum_{i=0}^{N_{b}} \left[ \frac{\partial^2 \mathcal{E}_b }{\partial \bar{\kappa}_{1,i} \partial \kappa_{1,i}} \left( \frac{\partial \kappa_{1,i} }{\partial \bar{\mathbf{q}}_{0} } \otimes  \frac{\partial \kappa_{1,i} }{\partial \bar{\mathbf{q}} } \right) \right] \\
    & +  \sum_{i=0}^{N_{b}} \left[  \frac{\partial^2 \mathcal{E}_b }{\partial \bar{\kappa}_{2,i} \partial \kappa_{2,i}} \left( \frac{\partial \kappa_{2,i} }{\partial \bar{\mathbf{q}}_{0} } \otimes  \frac{\partial \kappa_{2,i} }{\partial  \bar{\mathbf{q}} } \right) \right] \\
    \mathbb{J}_{t}^\mathrm{inv}  = \frac{\partial} {\partial \bar{\mathbf{q}}} \left( \frac{\partial \mathcal{E}_t}{\partial \mathbf{q}} \right) & =  \sum_{i=0}^{N_{b}} \left[ \frac{\partial^2 \mathcal{E}_t}{\partial \bar{\tau}_{i} \partial \tau_{i}} \left(\frac{\partial \tau_{i} }{\partial \bar{\mathbf{q}}_{0} } \otimes \frac{\partial \tau_{i} }{\partial   \bar{\mathbf{q}} } \right) \right].
\end{aligned}
\end{equation}
Note that the Hessian of the inverse-DER is non-symmetric.
Finally, the same Newton's iteration to enforce the residual error to be zero, and then we move to the next time step.

\begin{figure*}[h]  
\centering
\includegraphics[width=\textwidth]{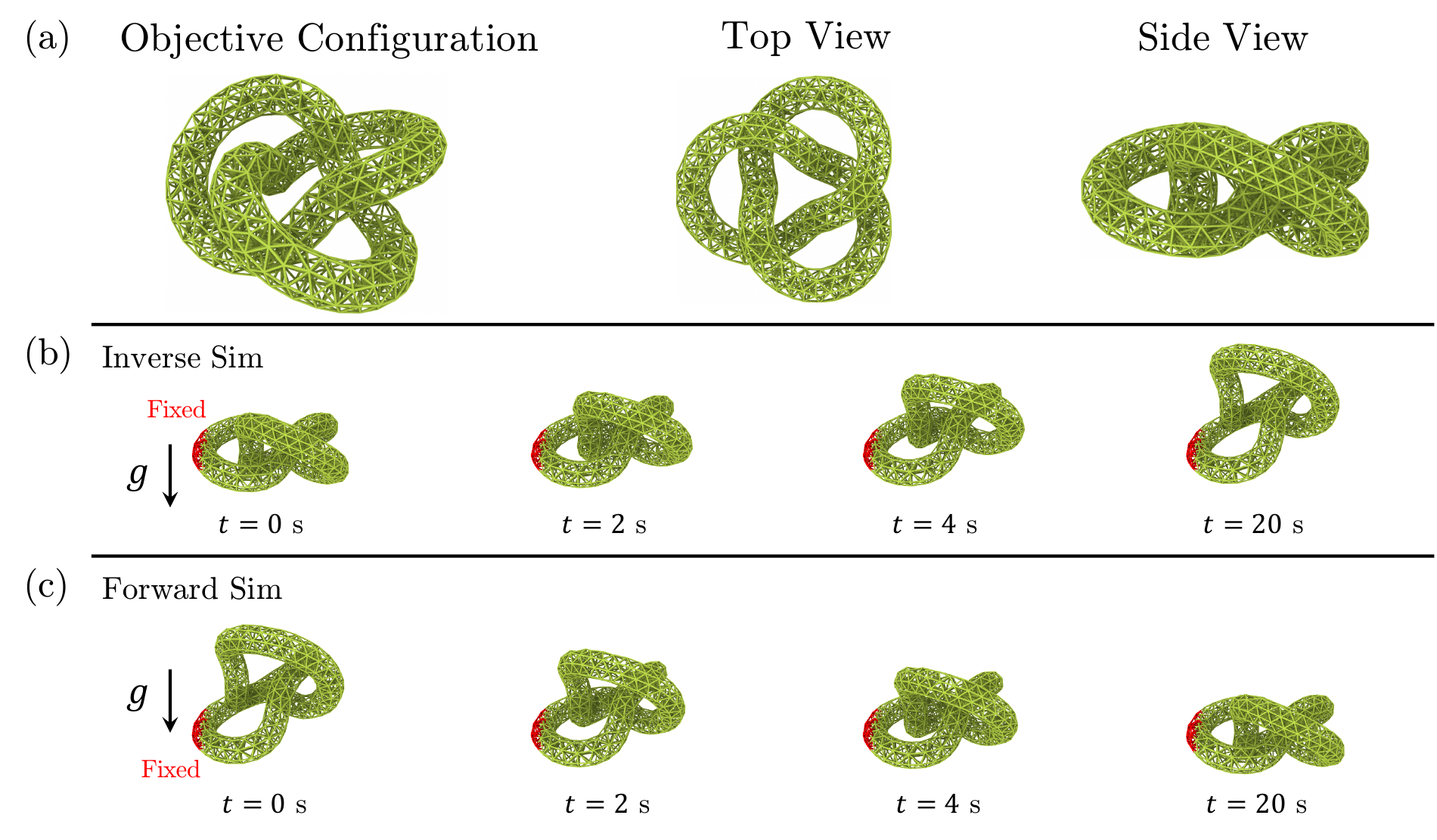}
\caption{\textbf{The inverse and forward simulation of net knot.} (a) The objective net knot. (b) The inverse simulation of net knot. (c) The verification of the inverse simulation results by forward simulation.}
\label{fig:case5}
\end{figure*}

\section{Results}
In this section, we present several inverse design examples using inverse-DER. Our approach enables direct computation of UC from DC. We emphasize two key innovations of this method. First, it effectively addresses inverse design problems for slender structures under general boundary conditions, including both clamped-free and clamped-clamped cases. Second, the framework is not limited to slender structures; it can be readily extended to net structures subjected to external loads, demonstrating strong potential for applications in architectural design. For all cases, the characteristic length of the rods is defined as their total length, while for net structures, it is defined as the diameter of the bounding sphere. The simulation parameters are listed in Table.~\ref{T1_1}.

\begin{table*}[t]
\centering
\normalsize
\setlength{\tabcolsep}{3.5pt} 
\begin{tabular}{lccccccc}
\toprule
Cases & Charac. length (m) & Radius (m) & Modulus (Pa) & Density (kg/m$^3$) & Gravity (m/s$^2$) & Magnetization (A/m) & Magnetic field (mT) \\
\midrule
Spherical curve             & 36.3   & 1e-2   & 1e7  & 1e3 & -             & -             & -         \\
Conical curve               & 7.0   & 1e-2   & 1e7  & 1e3 & -             & -             & -         \\
Hyperbolic curve            & 15.8   & 1e-2   & 1e7  & 1e3 & -             & -             & -         \\
Helix  (gravity)             & 25.4   & 1e-1   & 1e8  & 1e3 & [0,0,-10]     & -             & -         \\
Hyperbole (magnetic)            & 97.4   & 1e-1   & 1e8  & 1e3 & -             & [0,0,-1e5]    & [-5,0,0]  \\
Ring                        & 2.1 & 1.5e-3 & 5e6  & 2e2 & [0,0,-10]     & -             & -         \\
Knot                        & 6.4 & 1.5e-3 & 5e7  & 4e2 & [0,0,-10]     & -             & -         \\
Fulleren                    & 7.1   & 1e-2   & 6e6  & 5e2 & [0,0,-10]     & [0, 5e5, 0]   & [0,0,-1]  \\
\bottomrule
\end{tabular}
\caption{Simulation parameters for the different cases.}
\label{T1_1}
\end{table*}

\subsection{Single rod}

We show the inverse design of three types of curve-discretized surfaces with positive, zero and negative gaussian curvatures as shown in Fig.~\ref{fig:case123}. 
The equations of the spherical curve is:
\begin{equation}
    \begin{cases}
	x=\cos \left( 18\pi s \right) \sin \left( \pi s \right)\\
	y=\sin \left( 18\pi s \right) \sin \left( \pi s \right)\\
	z=\cos \left( \pi s \right)\\
\end{cases}\,\, \,\,\,\,\,\,\,\,                s\in \left[ 0,1 \right]
\end{equation}
The equations of the conical curve is:
\begin{equation}
    \begin{cases}
	x=(1-s/3)\cos(8\pi s/3)\\
	y=(1-s/3)\sin(8\pi s/3)\\
	z=2s/3\\
\end{cases}\,\,\,\,\,\,\,\,\,\,s\in \left[ 0,1 \right]
\end{equation}
The equations of hyperbolic curve is:
\begin{equation}
    \begin{cases}
	x=((s-1)^2+0.5)\cos(6\pi s) \,\,\\
	y=((s-1)^2+0.5)\sin(6\pi s)\\
	z=s/2\\
\end{cases}\,\,\,\,\,\,\,\,\,\,s\in \left[ 0.1,1.9 \right]
\end{equation}
For each curve, we aim to find the UD under clamped-clamped BCs.  


As illustrated in Fig.~\ref{fig:case1}, three representative target geometries are considered for the inverse design study: spherical, conical, and hyperbolic curves, each subjected to clamped–clamped boundary conditions. For each geometry, we first perform an inverse simulation to identify UCs corresponding to the DCs under compression. The resulting UCs are then validated through DER simulations for verification. To further assess the practical applicability of the proposed framework, the designed UCs are fabricated via 3D printing and experimentally stretched to reproduce the targets. As shown in Figs.~\ref{fig:case1} (a)–(c), both numerical simulations and experimental measurements exhibit excellent agreement with the target geometries, confirming that the recovered DCs closely match the design intent. These results collectively demonstrate the accuracy, reliability, and robustness of the proposed inverse-DER method for inverse design of slender structures.

\begin{figure}[h]  
\centering
\includegraphics[width=1\linewidth]{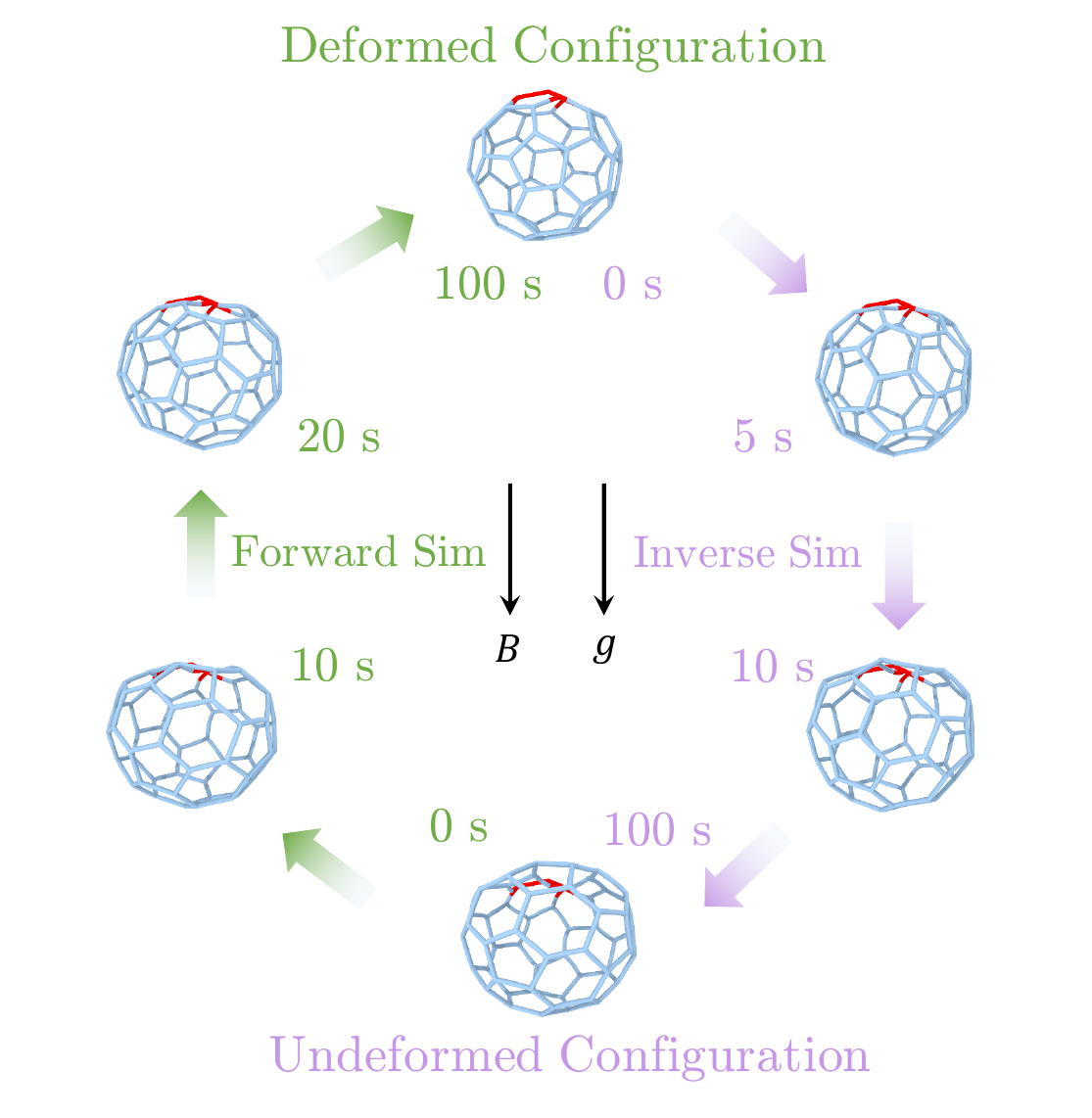}
\caption{The inverse and forward simulation of Fulleren under gravity and magnetic field.}
\label{fig:case6}
\end{figure}

We also consider the inverse design framework to slender structures subjected to clamped–free boundary conditions and external fields, such as gravity and magnetic fields, as illustrated in Fig.~\ref{fig:case23}. Specifically, we consider two representative cases: a helical curve exposed to varying gravitational accelerations, and a curve-discretized hyperbolic surface subjected to different magnetic fields. The parametric equation of the helical curve is given by:

\begin{equation}
\begin{cases}
x = 4s,\\
y = \sin (8\pi s),\\
z = \cos (8\pi s),
\end{cases}
\quad s \in [0, 1],
\end{equation}
and the equation of the curve-discretized hyperbolic surface is:
\begin{equation}
\begin{cases}
x = (4s^2 + 0.6)\cos(16\pi s),\\
y = (4s^2 + 0.6)\sin(16\pi s),\\
z = -4s,
\end{cases}
\quad s \in [0, 1].
\end{equation}
Figure~\ref{fig:case23} (a) illustrates the UCs of the helical curve obtained via inverse simulations under varying gravitational accelerations. These UCs are subsequently validated through DER simulations. Similarly, for the hyperbolic surface subjected to different magnetic field strengths, the inverse design procedure yields the corresponding UCs, as shown in Fig.~\ref{fig:case23} (b), which are also verified through DER simulations. The results demonstrate excellent agreement between the recovered UCs under external field loading and the target geometries, highlighting the predictive accuracy and robustness of the proposed inverse-DER method.

\subsection{Net Structures}
The primary advantage of incorporating stretch and bend elements lies in the inherent flexibility of our approach, which allows for a straightforward extension to complex net structures, as demonstrated in previous studies~\cite{huang2022nonlinear,huang2023numerical,huang2023contact}. By leveraging this capability, the inverse-DER method can be directly applied to determine UC of a wide range of net structures, bypassing the need for optimization. Moreover, the method’s adaptability enables systematic exploration of diverse structural topologies and mechanical responses, opening up new possibilities for the design of lightweight, compliant, and multifunctional architectures.

Although the identification of UC under gravitational loading has been extensively studied using asymptotic methods~\cite{chen2014asymptotic}, our inverse-DER framework provides several notable advantages. Unlike asymptotic techniques, which require laborious derivations of expansion coefficients and are often limited to simple geometries or loading conditions, the inverse-DER approach allows for the direct design of complex net structures subjected to external fields. Importantly, this is achieved while maintaining a computational cost comparable to that of a standard forward DER simulation. By circumventing the need for intricate asymptotic analysis, the proposed framework significantly streamlines the design process, broadens the range of applicable geometries, and facilitates the systematic exploration of multifunctional and adaptive structures under realistic loading scenarios.

The effectiveness of our inverse-DER approach is demonstrated through two representative examples: a twisting ring net (Fig.~\ref{fig:case4}) and a knot net (Fig.~\ref{fig:case5}). In both cases, the red segments are clamped while the remaining portions are free, imposing mixed boundary conditions that challenge traditional design methods. As shown in Fig.~\ref{fig:case4}, the UC of the ring net, obtained via the inverse-DER procedure, successfully reproduces the target configuration when subjected to forward simulation, exhibiting excellent agreement with the desired deformation. Similarly, the knot net (Fig.~\ref{fig:case5}) demonstrates consistent recovery of the intended shape, further validating the robustness and predictive accuracy of our framework. These examples highlight the capability of the proposed method to handle complex net topologies and mixed boundary conditions, offering a versatile and efficient tool for the inverse design of structurally intricate and mechanically responsive architectures.

\begin{table*}[t]
\begin{tabular}{cccccccc}
\hline
 & \#Vertices & \#Edges & \#Bends & Forward Time (s) & Forward Time/Step (ms) & Inverse Time (s) & Inverse Time/Step (ms) \\ \hline
Spherical curve & 500 & 499 & 498 & 1 & 6 & 1 & 7 \\ \hline
Conical curve & 500 & 499 & 498 & 1 & 6 & 1 & 7 \\ \hline
Hyperbolic curve & 500 & 499 & 498 & 1 & 7 & 1 & 6 \\ \hline
Ring net& 323 & 1552 & 14223 & 7 & 113 & 8 & 112 \\ \hline
Knot net& 600 & 2669 & 24872 & 13 & 219 & 14 & 208 \\ \hline
Fulleren & 510 & 540 & 630 & 3 & 6 & 3 & 7 \\ \hline
\end{tabular}
\caption{The comparison of simulation efficiency between forward and inverse simulation ran on a single core of an AMD Ryzen 7 $6800H$ @ $2.7$ GHz.}
\label{T2}
\end{table*}

Furthermore, the inverse-DER framework demonstrates strong capability in addressing multi-physical scenarios. As illustrated in Fig.~\ref{fig:case6}, we consider a magnetized Fullerene structure subjected simultaneously to gravitational and magnetic fields. The inverse simulation successfully find the UC, which is then rigorously validated through forward simulation. The excellent agreement between the DC and the target configuration confirms the predictive accuracy and robustness of the proposed method even under coupled multi-physical loading conditions. This example highlights the versatility of the inverse-DER approach, showing that it can systematically design complex structures that respond predictably to multiple external stimuli, thereby providing a powerful tool for the development of adaptive, multifunctional, and responsive materials.

\section{Discussions}

\subsection{Computational Efficiency}

The high computational efficiency is a very significant advantage of inverse-DER. 
In previous studies, inverse design is usually processed as an optimization problem, which leads to the greater consumption of computational resources than forward simulation. For example, the computational cost for solving the inverse problem with the LevMar solver, as reported by Chen et al., was about seventy times that of a single forward simulation~\cite{chen2014asymptotic}. However, because our inverse simulation is implemented within the Euler-Lagrange dynamics framework, the computational time required for inverse-DER is nearly identical to that of forward simulation as shown in Table.~\ref{T2},

\subsection{Existence of Inverse Solution}

Under some situations, UC does not exist under the specific BCs. We discuss the existence of the inverse solution in this section. For example, as shown in Fig.~\ref{fig:limitation}, we consider the DC as a horizontal cantilever beam and increase the elasto-gravitational parameter $\gamma$ gradually in inverse simulation. The elasto-gravitational parameter is defined as: $\gamma=\rho AgL^3/\mathrm{EI}_2$, where $A$ is the cross section, $\rho$ is the density, $L$ is the length of the beam, $g$ is the gravity parameter, $\mathrm{EI}_2$ is the bending stiffness. This 2D case can be solved analytically, as shown in Appendix B. We show the simulation results and theoretical result derived from inverse elastica theory (Appendix B) in Fig.~\ref{fig:limitation} (a). When the elasto-gravitational parameter is 9.04, the inverse simulation diverges. The physical meaning behind this phenomenon is that for a sufficiently flexible beam, we cannot find an UC to achieve the inverse design corresponding to DC under gravity. We also show the displacement of right end of UC as a function of $\gamma$ in Fig.~\ref{fig:limitation} (b) and the red vertical line $\gamma=3\pi$ is the maximum $\gamma$ from theoretical result.

\subsection{Energy Profiles of Forward and Inverse Simulation}

\begin{figure*}[h]  
\centering
\includegraphics[width=\textwidth]{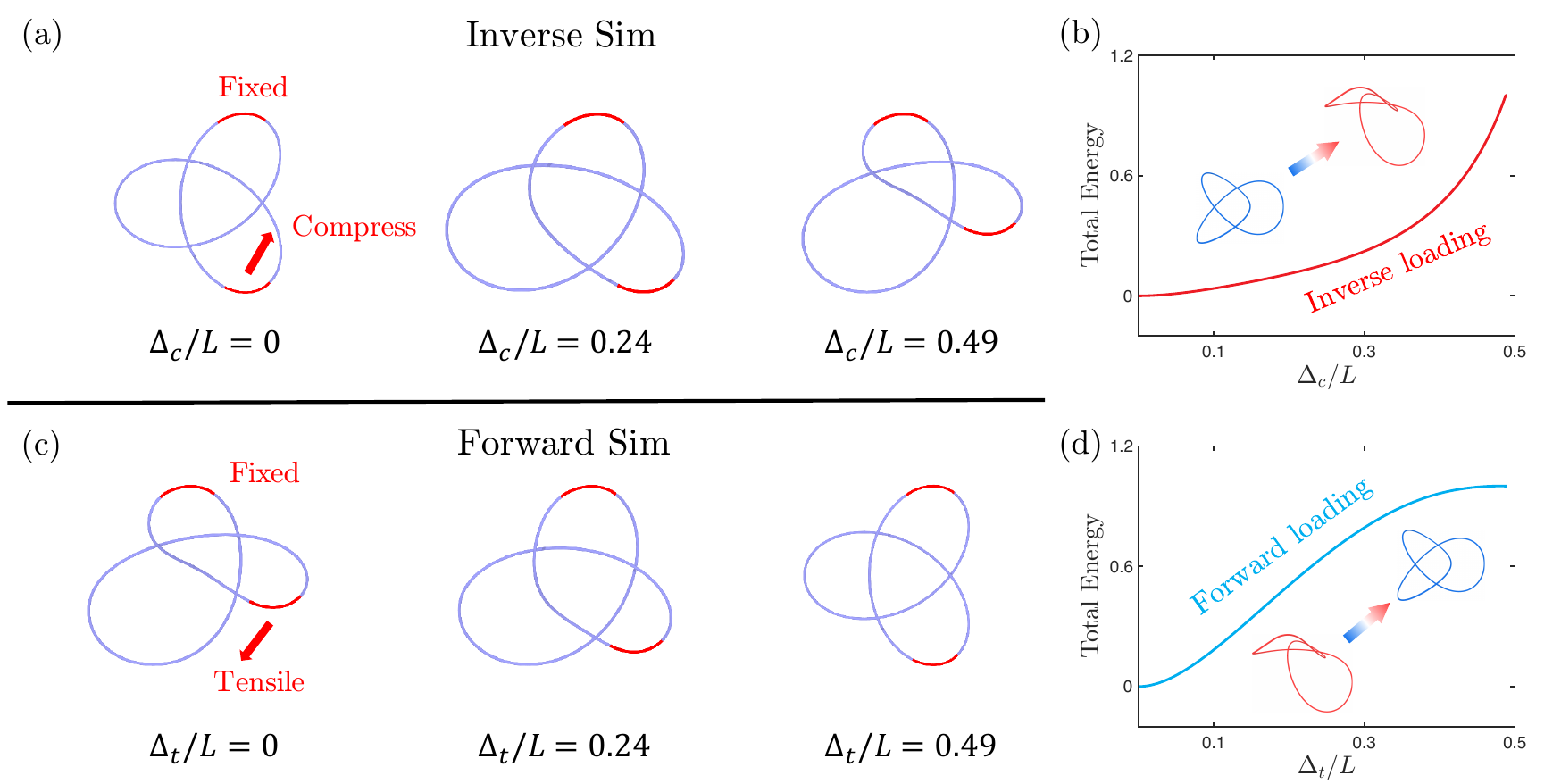}
\caption{\textbf{Inverse design of trefoil knot.} (a) Compression of a trefoil knot in inverse simulation, where $\Delta_c$ is the compression displacement and $L$ is the length of the trefoil knot. (b) Verification with the UC obtained from inverse simulation results, where $\Delta_t$ is the tensile displacement. (c) The normalized energy during inverse loading. (d) The normalized energy during forward loading.}
\label{fig:Knot}
\end{figure*}

We emphasize that inverse simulation is fundamentally distinct from the principle of time-reversal symmetry in classical mechanics. Note that systems with time-reversal symmetry exhibit identical energy trajectories under forward and time-reversed evolution, however, the energy profiles produced by forward and inverse simulation do not coincide. To illustrate this, we identify the UC of a compressed trefoil knot using inverse-DER. As shown in Fig.~\ref{fig:Knot}(a) and (b), a trefoil knot with fixed ends is compressed by a dimensionless displacement of 0.49 via inverse-DER, and the resulting UC is validated using forward DER. The corresponding energy profiles during inverse and forward simulation are presented in Fig.~\ref{fig:Knot}(c) and (d), respectively. The results confirm that although the total energy variation is identical in both simulations, their energy profiles are markedly different.

\subsection{Future Works}

In this section, we discuss the research directions for future work.

\textbf{Multiple Solutions and Bifurcation} For inverse design of an open elastic rod under gravity with clamped-free condition, we have demonstrated that the solution is unique.
However, for the more general boundary conditions, such as clamped-clamped boundary conditions, e.g., manipulation of a rod, there might be multiple solutions; thus, bifurcations may arise like the forward problem~\cite{yu2019bifurcations,yu2021numerical,huang2024exploiting}.
In the future, we will combine the arc-length method and stability analysis to further investigate the bifurcation effect existing in the inverse design problem in order to find all possible solutions.

\textbf{Extensible Inverse Simulation} Although this study focuses specifically on the inverse-DER, the proposed inverse simulation framework is readily extensible to any forward simulation algorithm based on the Euler–Lagrange formulation, such as discrete elastic shells, finite element methods, or peridynamics.
The central idea of our approach is to reformulate and solve the force equilibrium equations directly in the reference configuration, thereby addressing the inverse problem and facilitating the design of complex systems such as soft robots and mechanical metamaterials.

\textbf{Other External Field} We demonstrate inverse design here only under gravity and magnetic fields; however, the same approach can be applied to achieve inverse design of slender structures subjected to other types of external force fields, including fluid viscous forces, electric field forces, Coulomb interactions, van der Waals forces, and beyond.
All these interaction fields are conservative, and the potential energy differences can be computed directly from the initial and final configurations.
For non-conservative forces such as friction, the resulting force cannot be directly derived from only two configurations, as these systems exhibit path dependence. This category of problems will be investigated in our future work.

\begin{figure}[h]  
\centering
\includegraphics[width=\linewidth]{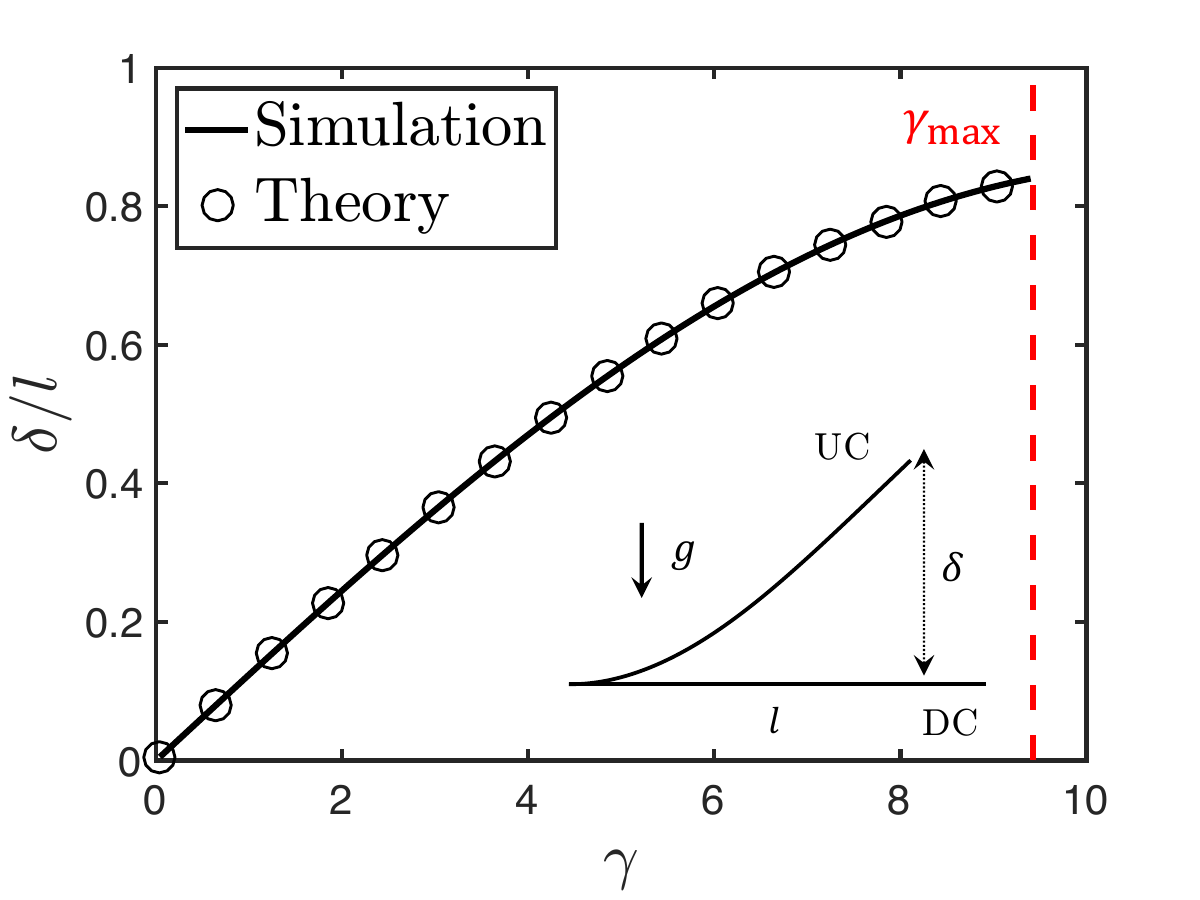}
\caption{\textbf{Inverse design of a cantilever beam.} The displacement of the right end for the inverse simulation of a cantilever beam under gravity. The solid line is the result of inverse-DER, while the circle is the theoretical solution.}
\label{fig:limitation}
\end{figure}

\section{Conclusion}

We build inverse-DER to address the inverse design of slender structures subject to general loading and boundary conditions.
The method is applied to recover UC under both displacement-controlled BCs and external physical fields.
Computationally, our approach maintains the same order of magnitude in cost as forward simulation.
We demonstrate the performance and robustness of the proposed framework through several inverse design examples on slender structures and nets.

While inverse-DER shows promising capability in solving the nonlinear systems arising in our inverse problems, several directions remain open for future work.
The method could be generalized to other comparable simulation frameworks.
For example, extending the approach from slender structures to 2D shells and 3D solids represents a natural next step.
Furthermore, the study of bifurcation behavior in inverse simulations presents an interesting research direction, as it may reveal deeper physical mechanisms.
Incorporating non-conservative forces, such as friction, also remains an important avenue for further investigation.



\bibliographystyle{ACM-Reference-Format}
\bibliography{publications}

\appendix

\section{Kirchhoff model and inverse Elastica}  

In this appendix, we briefly review the Kirchhoff model for both forward and inverse Elastica, which is a theoretical approach for a single rod-like slender system \cite{li2025inverse}.
The theoretical framework is developed for a single-rod system and is difficult to extend to arbitrary net-like configurations.
In the Kirchhoff model, the equilibrium equation of a slender structure can be written as: 
\begin{equation}
\begin{cases}
\mathbf{F}'+\mathbf{f }=0  \\
\mathbf{M}' + \mathbf{d}_3\times\mathbf{F} +\mathbf{m}= 0
\end{cases}
\end{equation}
where $(\cdot)'=\partial(\cdot)/\partial s$ is the derivative of variable $( \cdot )$ to arc length parameter $s$, $\mathbf{d}_3$ is the tangent vector, $\mathbf{F}$ is the internal force of the rod and $\mathbf{M}$ is the internal moment of the rod.  Later, we ignore the arc length parameter $(s)$ for simplification.
To solve the above ordinary differential equations (ODEs), we expand all physical vectors in the local deformed material frame (DC), $\mathbf{D}=\{\mathbf{d}_1,\mathbf{d}_2,\mathbf{d}_3\}^T$, (where $\mathbf{d}_1$ and $\mathbf{d}_2$ are the two material directors), the equilibrium equation and the linear constitutive equation are given as
\begin{equation}
\begin{aligned}
&\hat{\mathbf{F}}'+\hat{\mathbf{F}}\mathbf{\Omega}+\hat{\mathbf{f}}=0,\\
&\hat{\mathbf{M}}'+\hat{\mathbf{M}}\mathbf{\Omega }+\hat{\mathbf{F}}\mathbf{\Lambda }+\hat{\mathbf{m}}=0,\\
&\hat{\mathbf{M}}=\left( \mathbf{\omega }-\bar{\mathbf{\omega }} \right) \mathbf{S},\\
&\mathbf{\Lambda} =\left( \begin{matrix}
	0&		1&		0\\
	-1&		0&		0\\
	0&		0&		0\\
\end{matrix} \right), 
\end{aligned}
\end{equation}
where $\mathbf{\Omega} = \rm{skew} (\mathbf{\omega})$ is the rotation tensor, $\mathbf{\omega}=\{\omega_1,\omega_2,\omega_3\}$ is the Darboux vector of DC, $\bar{\mathbf{\omega}}$ is the Darboux vector of UC, $\mathbf{S}=\rm{diag}(\{EI_1,EI_2,GJ\})$ is the stiffness matrix. Also, the physical vectors for the force and moment are transformed into the component vector, i.e., $\hat{\mathbf{F}}=\{F_1,F_2,F_3\}$ is the component vector of internal force vector $\mathbf{F}$ in deformed material frame thus $\mathbf{F}=\hat{\mathbf{F}}{\mathbf{D}}$, and, similar for $\hat{\mathbf{M}}=\{M_1,M_2,M_3\}$ (and $\mathbf{M}=\hat{\mathbf{M}}\mathbf{D}$),
$\hat{\mathbf{f}}=\{f_1,f_2,f_3\}$ (and $\mathbf{f}=\hat{\mathbf{f}}\mathbf{D}$),
and $\hat{\mathbf{m}}=\{m_1,m_2,m_3\}$ (and $\mathbf{m}=\hat{\mathbf{m}}\mathbf{D}$). 

\paragraph{Kirchhoff model}
For the Kirchhoff model, we need to consider the geometric constraints for DC:
\begin{equation}
\begin{cases}
\mathbf{\Gamma }'=\mathbf{R}_{3}^{T}, \\
\mathbf{q}'=\mathbf{Q}\mathbf{q}.
\end{cases}
\label{eqn:12}
\end{equation}

where $\mathbf{\Gamma}=\{x(s),y(s),z(s)\}^{T}$ is coordination of center line of DC as a function of arc length parameter $s$, $\mathbf{q}=\{q_{0}(s),q_{1}(s),q_{2}(s),q_{3}(s)\}^T$ is the quaternion vector of DC, $\mathbf{R}_{3}$ is the third row of rotation matrix $\mathbf{R}$ and $\mathbf{Q}$ is a quaternion matrix for DC. 
$\mathbf{R}$ and $\mathbf{Q}$ can be written as:
\begin{equation}
\mathbf{R}=2\left( \begin{matrix}
	q_{0}^{2}+q_{1}^{2}-1/2&		q_{1}q_{2}+q_{0}q_{3}&		q_{1}q_{3}-q_{0}q_{2}\\
	q_{1}q_{2}-q_{0}q_{3}&		q_{0}^{2}+q_{2}^{2}-1/2&		q_{2}q_{3}+q_{0}q_{1}\\
	q_{1}q_{3}+q_{0}q_{2}&		q_{2}q_{3}-q_{0}q_{1}&		q_{0}^{2}+q_{3}^{2}-1/2\\
\end{matrix} \right)
    \label{eqn:30}
\end{equation}
and
\begin{equation}
    \mathbf{Q}=\left( \begin{matrix}
	0&		-\omega\\
	\omega^T&		\Omega\\
\end{matrix} \right). 
    \label{eqn:31}
\end{equation}
The Kirchoff model can be summarized as:
\begin{equation}
 \begin{cases}
\hat{\mathbf{F}}' +\hat{\mathbf{F}}\mathbf{\Omega }+\hat{\mathbf{f}}=0
\\
\hat{\mathbf{M}}' +\hat{\mathbf{M}}\mathbf{\Omega }+\hat{\mathbf{F}}\mathbf{\Lambda }+\hat{\mathbf{m}}=0
\\
\hat{\mathbf{M}}=\left( \mathbf{\omega }-\bar{\mathbf{\omega}} \right) \mathbf{S}
\\
\mathbf{\Gamma}' =\mathbf{R}_{3}^{T}
\\
\mathbf{q}' =\mathbf{Q}\mathbf{q}
\end{cases},
\label{eqn:Kirchhoff}
\end{equation}
In Eq.~\eqref{eqn:Kirchhoff}, the unknowns are the internal force: $\hat{\mathbf{F}}$, $\hat{\mathbf{M}}$, and the geometry of the DC: $\mathbf{\omega}$,  $\mathbf{\Gamma}$, and $\mathbf{q}$.
\paragraph{Inverse Elastica}
The most significant in inverse elastica theory is that the geometry equation of UC is introduced:
\begin{equation}
\begin{cases}
\bar{\mathbf{\Gamma }}'=\bar{\mathbf{R}}_{3}^{T},\\
\bar{\mathbf{q}}'=\bar{\mathbf{Q}}\bar{\mathbf{q}}.
\end{cases}
\label{eqn:13}
\end{equation}
where $\bar{\mathbf{\Gamma}}=\{\bar{x}(s),\bar{y}(s),\bar{z}(s)\}^{T}$ is coordination of center line of UC as a function of arc length parameter $s$, $\bar{\mathbf{q}}=\{\bar{q}_{0}(s),\bar{q}_{1}(s),\bar{q}_{2}(s),\bar{q}_{3}(s)\}^T$ is the quaternion vector of UC, $\bar{\mathbf{R}}_{3}$ is the third row of rotation matrix $\bar{\mathbf{R}}$ and $\bar{\mathbf{\rm{Q}}}$ is a quaternion matrix for UC similar to Eq.~\eqref{eqn:30} and~\eqref{eqn:31}.

Now we derive the complete inverse elastica theory:
\begin{equation}
\begin{cases}
\hat{\mathbf{F}}' +\hat{\mathbf{F}}\mathbf{\Omega }+\hat{\mathbf{f}}=0
\\
\hat{\mathbf{M}}' +\hat{\mathbf{M}}\mathbf{\Omega }+\hat{\mathbf{F}}\mathbf{\Lambda }+\hat{\mathbf{m}}=0
\\
\hat{\mathbf{M}}=\left( \mathbf{\omega }-\bar{\mathbf{\omega}} \right) \mathbf{S}
\\
\bar{\mathbf{\Gamma}}' =\bar{\mathbf{R}}_{3}^{T}
\\
\bar{\mathbf{q}}' =\bar{\mathbf{Q}}\bar{\mathbf{q}}
\end{cases}
    \label{eqn:inverse}
\end{equation}
In Eq.~\eqref{eqn:inverse}, the unknowns are $\bar{\mathbf{\omega}}$, $\bar{\mathbf{\Gamma}}$, $\hat{\mathbf{F}}$, $\hat{\mathbf{M}}$, and $\bar{\mathbf{q}}$. These variables can be determined from the prescribed displacement boundary conditions $\bar{\mathbf{q}}(0)$, $\bar{\mathbf{q}}(L)$, $\bar{\mathbf{\Gamma}}(0)$, and $\bar{\mathbf{\Gamma}}(L)$ for UC, indicating that the inverse problem can be solved directly in a manner analogous to the forward problem. 
If we interpret $\mathbf{\omega}$ as the Darboux vector of the reference configuration (DC) and $\bar{\mathbf{\Gamma}}$ as the centerline of the current configuration (UC) to be solved, Eq.~\eqref{eqn:inverse} bears a strong resemblance to the Kirchhoff model. The key distinction is that the forces and moments in Eq.~\eqref{eqn:inverse} are expressed in the material frame of the reference configuration (DC) rather than in that of the current configuration (UC), which is within our framework in section~4.1.

\section{Analytical solution for inverse design of cantilever beam}
We consider the 2D case, Eq.~\eqref{eqn:inverse} can be simplified as:
\begin{equation}
    \begin{cases}
	 F_1'=-F_3\omega _2+f_1\\
	F_3'=F_1\omega _2+f_3\\
	M_2'=-F_1+m_2\\
       M_2 = \mathrm{EI}_2(\omega_2-\bar{\omega}_2) \\
\end{cases}
    \label{eqn:2d}
\end{equation}
Here, we introduce the rotation angle $\bar{\theta}$, which is the rotation angle of UC ($\bar{\omega}_2=\bar{\theta}'$). When we consider a cantilever beam as a target under gravity, we have $\omega_2=0$, $f_3=0$, $f_1=\rho A g$, where $g$ is the gravity, $\rho$ is the density of the beam, $A$ is the cross section. Then Eq.~\eqref{eqn:2d} can be simplified as:
\begin{equation}
     \mathrm{EI}_2\bar{\theta}''=\rho Ag(s-L)
     \label{eqn:2d_beam}
\end{equation}
Note that for cantilever beam, it meets $M_2=0$ at the free end. Thus we have the boundary condition for Eq.~\eqref{eqn:2d_beam}:  $\bar{\theta}(0)=0$ and $\bar{\theta}'(L)=0$. Finally we can solve $\bar{\theta}={\gamma}\widetilde{s}(\widetilde{s}^2-3\widetilde{s}+3)/6$, where we have use the dimensionless parameter $\gamma={\rho Ag L^3}/{\mathrm{EI}_2}$ and $\widetilde{s}=s/L$, $L$ is the length of the beam. The shape of the beam can be solved by $dx/ds=\cos\theta$ and $dy/ds=\sin\theta$, which is the theoretical solution in Fig.~\ref{fig:limitation} (a). 
We consider a extreme case, the rotation angle at $s=L$ is $\pi/2$, which means the deformation between UC and DC is maximum. We can solve the maximum $\gamma=3\pi$  theoretically.



\section{Code}

The implementation of inverse-DER is available in our public GitHub repository: \url{https://github.com/weicheng-huang-mechanics/inverse_rod_simulation}

\section{Video}

We provide a video to demonstrate our inverse-DER numerical framework.

\end{document}